\documentclass[12pt]{iopart}   
\usepackage{graphicx}
\usepackage{epsfig}
\usepackage{amssymb}
\begin{document}

\newcommand{\be}{\begin{equation}}
\newcommand{\ee}{\end{equation}}
\newcommand{\beqn}{\begin{eqnarray}} 
\newcommand{\eeqn}{\end{eqnarray}}

\title[Critical behavior of models with infinite disorder at a star junction of chains]{Critical behavior of models with infinite disorder at a star junction of chains}

\author{R\'obert Juh\'asz}
\address{Institute for Solid
State Physics and Optics, Wigner Research Centre for Physics, H-1525 Budapest,
P.O. Box 49, Hungary}
\ead{juhasz.robert@wigner.mta.hu}

\begin{abstract}
We study two models having an infinite-disorder critical point --- the zero temperature random transverse-field Ising model and the random contact process --- on a star-like network composed of $M$ semi-infinite chains connected to a common central site. 
By the strong disorder renormalization group method, the scaling dimension $x_M$ of the local order parameter at the junction is calculated. 
It is found to decrease rapidly with the number $M$ of arms, but remains positive for any finite $M$. This means that, in contrast with the  pure 
transverse-field Ising model, where the transition becomes of first order for $M>2$, it remains continuous in the disordered models, although, for not too small $M$, it is hardly distinguishable from a discontinuous one owing to a close-to-zero $x_M$. 
The scaling behavior of the order parameter in the Griffiths-McCoy phase is also analyzed. 
\end{abstract}

\maketitle

\section{Introduction}

The description of phase transitions in equilibrium or nonequilibrium many-body systems in the presence of quenched disorder is a challenging problem with many unresolved questions \cite{im,vojta_rev}. 
Models with quenched disorder are usually harder to treat than their pure counterparts, with an important exception of strongly disordered systems, in which the (initially finite) variations of the strength of disorder are growing without limits as the system is coarse-grained and determine the large-scale behavior in critical points \cite{fisher,im}. 
One of the most prominent representatives of this class of models is the random transverse-field Ising model (RTIM), which exhibits ferromagnetic order at zero temperature below a critical strength of the transverse field. 
The quantum critical point of the model is described by an {\it infinite-disorder fixed point} (IDFP) of the above mentioned coarse-graining procedure known as {\it strong disorder renormalization group} (SDRG) method \cite{fisher,mdh}. 
Another representative is the disordered variant of the {\it contact process} \cite{cp,liggett}, which is a paradigmatic model in the field of population dynamics and epidemics. The contact process is a stochastic model consisting of reproduction and death of individuals living on a lattice and possesses an absorbing phase transition from a fluctuating phase to an absorbing one \cite{hhl}. 
Although the phase transitions in the pure variants of these models belong to different universality classes \cite{odor}, in the presence of disorder, they turn out to share the same universal critical exponents as it has been predicted by an SDRG approach of the one dimensional contact process \cite{hiv}, at least for strong enough initial strength of the disorder. Whether any weak initial disorder leads to an IDFP critical behavior of the contact process (which is the case for the RTIM) is an open question but this scenario has been confirmed in one dimension for a relatively weak initial disorder \cite{vd}.
 
If the underlying lattice is one dimensional, it remains so under the SDRG procedure and, in addition to this, the parameters on different sites remain independent (if they were so initially), which allows for an analytical solution of the SDRG flow equations and the determination of the critical exponents 
\cite{fisher}.   
Unfortunately, this is restricted to one-dimension; otherwise one resorts to a numerical implementation of the SDRG scheme and an estimation of the critical exponents \cite{ki_d}. 
Thinking of the interpretations of the contact process, a realistic modeling would require a study of the model in two dimensions (typically for population dynamics) or on complex networks (for epidemic spreading)  
but in these cases, the analytical tractability is lost.  
Keeping solvability, we will make one step toward more complex topologies in this work and study the above models on a star-like network that is composed of $M$ semi-infinite chains connected to a central site.
This arrangement may be a toy model for describing the local critical behavior near a junction in a network that consists of long one-dimensional segments with junctions far away from each other. 

Note that the semi-infinite chain and the infinite chain are special cases of this network ($M=1$ and $M=2$, respectively).  
The local critical behavior of the RTIM has been studied in both cases and the scaling exponent of the surface order parameter at the end point of a semi-infinite chain has been found to be different from that characterizing the bulk 
 order parameter \cite{mccoy,fisher}. 

The {\it pure} variant of the transverse-field Ising model at multiple junctions has been studied in earlier works from different aspects. 
The criticality condition has been rigorously shown to be identical to that of the one-dimensional model \cite{bjornberg}. 
With a somewhat different type of junction, namely, when the first spins of chains are coupled with those of all other chains, the scaling dimension of the magnetization near the junction has been shown to be $1/2$ if $M<1$, coupling-dependent for $M=2$, and zero for $M>2$ \cite{itb}. The latter means that the transition at the junction is of first order for $M>2$, having a non-zero spontaneous magnetization. The limit $M=0$ provides the surface magnetization exponent of the classical two-dimensional Ising model in the presence of a random surface field. 
In addition to this, Majorana fermions and the topological Kondo effect have also been recently studied in this model \cite{tsvelik}.  

Due to the one-dimensional character of the network apart from the center, 
certain questions can be treated analytically by the SDRG method but still the local critical behavior near the center will turn out to be non-trivial and characterized by a different scaling exponent for each value of the number $M$ of the arms of the network.
We will show that, after solving the appropriate SDRG flow equation formulated for the survival probability of the central spin, the scaling exponent of the order parameter at the center can be obtained as the smallest eigenvalue of a tridiagonal matrix of order $M$.  
It is found to decrease rapidly to zero with increasing $M$ but remains positive for any finite $M$. 
 
The paper is organized as follows. In section \ref{models}, the definitions of models are given and their SDRG schemes are reviewed. 
In section \ref{sdrg}, the survival probability of the central site in the SDRG procedure is calculated both in the critical point and in the disordered phase. 
Relations to the critical exponents characterizing the physical quantities of the models are provided by means of a scaling theory in section \ref{scaling}. 
Finally, the results are discussed in section \ref{discussion}.

\section{Models and their SDRG approaches}
\label{models}

\subsection{The random transverse-field Ising chain} 

The random transverse-field Ising chain (RTIC) is defined by the Hamiltonian 
\be 
H=-\sum_nJ_n\sigma_n^x\sigma_{n+1}^x -\sum_nh_n\sigma_n^z,
\label{rtic}
\ee
where $\sigma_n^x$ and $\sigma_n^z$ are Pauli operators on site $n$, and 
the bonds $J_n$ and transverse fields $h_n$ are positive, i.i.d. quenched 
random variables.
At zero temperature, the model undergoes a quantum phase transition from a ferromagnetic phase to a paramagnetic one as the quantum control parameter 
$\Delta=\overline{\ln J}-\overline{\ln h}$ is varied.  Here and in the followings, the overbar denotes an average over the distribution of parameters.
In the ferromagnetic phase ($\Delta>0$), the average spontaneous magnetization $\overline{m}_0=\lim_{H\to 0}\overline{\langle\sigma_n^x\rangle}(H)$ in response to a vanishing longitudinal magnetic field $H$ applied on spin $n$ only is positive and vanishes as 
\be 
\overline{m}_0(\Delta)\sim \Delta^{\beta}
\ee 
as the critical point at $\Delta=0$ is approached. 

In the SDRG treatment of the RTIC formulated by Fisher \cite{fisher},  the energy scale $\Omega=\max_n\{J_n,h_n\}$ is gradually reduced by iteratively applying two kinds of decimation steps, in which the effective couplings are obtained by a perturbation calculation.  
First, if the largest coupling is a field $\Omega=h_n$, furthermore $h_n\gg J_{n-1},J_n$, the spin on site $n$, being pinned in the direction of the transverse field, is decimated and spins on site $n-1$ and $n+1$ are connected directly by an effective bond 
\be 
\tilde J=J_{n-1}J_n/h_n.
\label{hdec}
\ee
Second, if the largest coupling is a bond $\Omega=J_n\gg h_{n},h_{n+1}$, 
the spins $n$ and $n+1$ form a giant spin, which experiences an effective field 
\be 
\tilde h=h_nh_{n+1}/J_n.
\label{Jdec}
\ee 
Since the distributions of logarithmic couplings broaden without limits during the procedure, the above perturbative reduction steps become more and more accurate and asymptotically exact in the critical fixed point. 
As the parameters on different sites remain independent throughout the SDRG procedure, it is sufficient to deal with their distributions 
$R_{\Gamma}(\beta)$ and $P_{\Gamma}(\zeta)$, where $\beta=\ln(\Omega/h)$, 
$\zeta=\ln(\Omega/J)$ and $\Gamma=\ln(\Omega_0/\Omega)$ are logarithmic variables and $\Omega_0$ denotes the initial value of the energy scale. 
Note that, in terms of these variables, the decimation rules in Eqs. (\ref{hdec}) and (\ref{Jdec}) are simply $\tilde\zeta=\zeta_{n-1}+\zeta_n$ and $\tilde\beta=\beta_{n-1}+\beta_n$.
Their evolution under the change of $\Gamma$ is described by the equations 
\beqn 
\frac{\partial R_{\Gamma}(\beta)}{\partial\Gamma}=
\frac{\partial R_{\Gamma}(\beta)}{\partial\beta} + p_0(\Gamma)\int d\beta'R_{\Gamma}(\beta')R_{\Gamma}(\beta-\beta') +  R_{\Gamma}(\beta)[r_0(\Gamma)-p_0(\Gamma)] 
\nonumber \\
\frac{\partial P_{\Gamma}(\zeta)}{\partial\Gamma}=
\frac{\partial P_{\Gamma}(\zeta)}{\partial\zeta} + r_0(\Gamma)\int d\zeta'P_{\Gamma}(\zeta')P_{\Gamma}(\zeta-\zeta') +  P_{\Gamma}(\zeta)[p_0(\Gamma)-r_0(\Gamma)],  
\label{rgflow}
\eeqn
where $p_0(\Gamma)=P_{\Gamma}(0)$ and $r_0(\Gamma)=R_{\Gamma}(0)$ \cite{fisher}. 
The fixed-point solutions, which are attractors for any sufficiently regular initial distributions, are of the form \cite{fisher}
\be 
R_{\Gamma}(\beta)=r_0(\Gamma)e^{-r_0(\Gamma)\beta}, \qquad 
P_{\Gamma}(\zeta)=p_0(\Gamma)e^{-p_0(\Gamma)\zeta}
\label{RP}
\ee
and the functions  $p_0(\Gamma)$ and $r_0(\Gamma)$ obey the differential equations 
\be 
\frac{dp_0(\Gamma)}{d\Gamma}=\frac{dr_0(\Gamma)}{d\Gamma}=-p_0(\Gamma)r_0(\Gamma).
\label{pr}
\ee 
In the critical fixed point 
\be 
p_0(\Gamma)=r_0(\Gamma)=\frac{1}{\Gamma+\Gamma_0},
\label{prcrit}
\ee
where the constant $\Gamma_0$ characterizes the initial strength of the 
disorder.
Here, the relationship between the length scale $\xi$ (the inverse of the fraction $n$ of non-decimated spins) and the logarithmic energy scale is given by the solution of the equation $dn/d\Gamma=-n(p_0+r_0)$ in the form 
\be 
\Gamma\sim \xi^{\psi}
\label{dyn}
\ee
with $\psi=1/2$. 

On the paramagnetic side of the critical point ($\Delta<0$), where the Griffiths-McCoy phase takes place \cite{griffiths}, the SDRG transformation has a line of fixed points parameterized by the disorder-dependent dynamical exponent $z$, which appears in the scale factors as 
\be 
p_0(\Gamma)=r_0(\Gamma)-\frac{1}{z}=\frac{Ce^{-\Gamma/z}}{1-Ce^{-\Gamma/z}}=
O(e^{-\Gamma/z}).
\label{prgp}
\ee  
Here, $C$ is a non-universal constant. 
The energy-length relationship is $\Omega\sim\xi^z$ in this phase, and $z$ is related to the initial distribution of parameters through the implicit equation \cite{igloi}
\be 
\overline{\left(\frac{J}{h}\right)^{1/z}}=1.
\ee

\subsection{The disordered contact process} 

The contact process is a continuous-time stochastic process on a lattice, the sites of which can be either active or inactive. 
In its one-dimensional, disordered variant, 
the following two kinds of transitions can occur independently. 
First, an active site $n$ activates its neighboring sites $n-1$ and $n+1$ with 
rates $\lambda_{n-1}$ and $\lambda_n$, respectively. 
Second, an active site $n$ becomes inactive with a rate $\mu_n$. 
The transition rates $\lambda_n$ and $\mu_n$ are assumed to be 
i.i.d. quenched random variables. 
The fraction $\overline{\rho}$ of active sites in the steady state, which is positive in the active phase, vanishes continuously as the control parameter 
$\Delta=\overline{\ln\lambda}-\overline{\ln\mu}$ approaches a critical value $\Delta_c$ as $\overline{\rho}(\Delta)\sim (\Delta-\Delta_c)^{\beta}$. 

The SDRG scheme of the model is formally similar to that of the RTIC \cite{hiv}.
The rate scale $\Omega=\max_n\{\lambda_n,\mu_n\}$ is gradually reduced by two kinds of perturbative decimation steps. 
If the largest rate is $\mu_n$, site $n$ is decimated and the neighboring sites are connected by a link with an effective activation rate $\tilde\lambda=\lambda_{n-1}\lambda_n/\mu_n$. If $\Omega=\lambda_n$, site $n$ and $n+1$ are merged to a cluster having an effective rate $\tilde\mu_n=2\mu_n\mu_{n+1}/\lambda_n$.
These decimation rules are identical to those of the RTIC with the correspondences $\lambda\leftrightarrow J$, $\mu\leftrightarrow h$ apart from a factor of $2$, which is, however, irrelevant in the critical fixed point. 
As a consequence, the critical exponents such as $\beta$ of the two model are identical. 
   
\section{SDRG scheme on a star junction of chains}
\label{sdrg}

After reviewing the SDRG technique in one dimension, let us turn to the RTIM and the contact process on a star-like network comprising $M$ chains. 
The former model is composed of semi-infinite RTICs labeled by $p=1,2,\dots,M$
\be 
H_p=\sum_{n=1}^{\infty}J_{n,p}\sigma_{n,p}^x\sigma_{n+1,p}^x+\sum_{n=1}^{\infty}h_{n,p}\sigma_{n,p}^z,
\ee
the first spins of which are coupled to a central spin labeled by $0$, so that the Hamiltonian of the total system is 
\be 
H_M=\sum_{p=1}^MH_p+H_c,
\ee
where 
\be 
H_c=\sum_{p=1}^MJ_{0,p}\sigma_0^x\sigma_{1,p}^x + h_0\sigma_0^z.
\ee
For the sake of simplicity, the bonds $J_{0,p}$ and the field $h_0$ are drawn from the same distributions as the bulk parameters. 

The contact process on a star-like network is defined in the arms as before while, at the junction, the dynamical rules are given as follows.
If the first site of arm $p$ is active, it activates the central site with a rate $\lambda_{0,p}$ provided the latter is inactive, and this occurs vice versa, as well, with the same rate. Furthermore, if the central site is active it becomes inactive with a rate $\mu_0$. The rates $\lambda_{0,p}$ and $\mu_0$ are drawn from the same distributions as the bulk rates.

As long as the central spin is not decimated in the SDRG procedure, the decimation rules will be identical to those formulated in a one-dimensional chain and the topology of the network will remain unaltered. 
If, however, the field of the central spin is decimated, then new bonds between all pairs $(p,q)$ of its $M$ neighboring spins are created, each according to the one-dimensional rule in Eq. (\ref{hdec}), i.e.
\be 
\tilde J_{p,q}=J_{0,p}J_{0,q}/h_0. 
%\qquad p,q=1,2,\dots, M.
\ee
In this way, a complete graph with $M$ sites forms in the center, with semi-infinite chains attached to them. 
If, in a later stadium of the SDRG procedure, a bond of this complete graph is decimated, its size reduces from $M$ to $M-1$, so that, during the renormalization, the size of the complete graph in the center fluctuates between $1$ and $M$. 
In the particular case of $M=3$, the center can be found in one of two configurations: a delta junction or a star junction, namely.   

Since the bonds in the central part are correlated with each other and, after field or bond decimations, double bonds between clusters appear, which are to be added,  
it is difficult to treat analytically the evolution of the central part of the system under the renormalization even for $M=3$. Nevertheless, if one is interested in the scaling of the {\it average} order parameter, it is sufficient to deal with a much simpler quantity, the probability $S(\Gamma)$, namely, that the central spin is still active in a cluster (i.e. not decimated) at the logarithmic energy scale $\Gamma$. 
As we will discuss in section \ref{scaling}, the dependence of the order parameter on physical quantities like the strength of an external field, the control parameter or, in the case of the contact process, dependence on time in non-stationary states, can be derived from the properties of this function by scaling considerations. 

Thus, the first task is to determine the probability $S(\Gamma)$ for a general $M$. 
This can be achieved by writing and solving an evolution equation for 
the probability $s_{\Gamma}(\beta)d\beta$ that the central spin is active at scale $\Gamma$ in a cluster having an effective (logarithmic) field $\beta$. 
In the way analogous to that leading to the SDRG flow equations (\ref{rgflow}), one can formulate the following evolution equation for $s_{\Gamma}(\beta)$ under the change of the energy scale $\Gamma$: 
\be 
\frac{\partial s_{\Gamma}(\beta)}{\partial\Gamma}=
\frac{\partial s_{\Gamma}(\beta)}{\partial\beta} - Mp_0(\Gamma)
\left[s_{\Gamma}(\beta) - \int_0^{\beta}d\beta's_{\Gamma}(\beta')R_{\Gamma}(\beta-\beta')\right].
\label{sflow}
\ee
The first term on the r.h.s. appears owing to the dependence of $\beta$ on $\Gamma$. The second term describes the change of the field of the central cluster when any of the first bonds in the arms is decimated and the first cluster in that arm is merged with the central one. 
These events occur in the arms independently, hence the factor $M$. 
When formulating Eq. (\ref{sflow}), it has been used that, due to the independence of couplings on different places, the distribution of the first bond and the first field in any arm under the condition that the central cluster has not yet been decimated is identical to the distribution $P_{\Gamma}(\zeta)$ and $R_{\Gamma}(\beta)$, respectively, of an infinite chain.   
Eq. (\ref{sflow}) has been formulated and solved for the special case of a semi-infinite chain ($M=1$) by Fisher \cite{fisher} and for an infinite chain ($M=2$) by Refael and Moore \cite{rm} in the critical point. 

The solution of Eq. (\ref{sflow}) for a general $M$ can be found by the ansatz 
\be 
s_{\Gamma}(\beta)=\left\{\sum_{n=0}^{M-1}a_n(\Gamma)[r_0(\Gamma)\beta]^n\right\}
r_0(\Gamma)e^{-r_0(\Gamma)\beta}, 
\label{ansatz}
\ee
which contains the unknown functions $a_n(\Gamma)$ $n=0,1,2,\dots,M-1$ of $\Gamma$. 
The probability $S(\Gamma)$ we are looking for is related to these functions as follows: 
\be 
S(\Gamma)=\int_0^{\infty}d\beta s_{\Gamma}(\beta)=
\sum_{n=0}^{M-1}a_n(\Gamma)\int_0^{\infty}d\eta e^{-\eta}\eta^n= 
\sum_{n=0}^{M-1}(n!)a_n(\Gamma).
\label{Sgamma}
\ee
Here, we have introduced the variable $\eta\equiv r_0(\Gamma)\beta$.
Substituting the expression in Eq. (\ref{ansatz}) into Eq. (\ref{sflow}) and using the relations (\ref{pr}) in order to eliminate the derivative of $r_0(\Gamma)$, one obtains after a lengthy calculation  
\beqn 
\sum_{n=0}^{M-1}\frac{da_n}{d\Gamma}\eta^n=p_0\sum_{n=0}^{M-1}na_n\eta^n + 
\sum_{n=0}^{M-1}a_n\eta^n[(1-M)p_0-r_0] - p_0\sum_{n=0}^{M-1}a_n\eta^{n+1}+ 
\nonumber \\
+r_0\sum_{n=1}^{M-1}na_n\eta^{n-1}+Mp_0\sum_{n=1}^{M-1}\frac{a_n}{n+1}\eta^{n+1}. 
\eeqn
The ansatz (\ref{ansatz}) will be a solution if this equality holds for all $\eta$, which requires the coefficients of all powers of $\eta$ to be identical on the two sides.  
The terms proportional to $\eta^M$ cancel while, from the comparison of the lower order terms, we obtain the following differential equations for the unknown functions $a_n(\Gamma)$: 
\be 
\frac{da_n}{d\Gamma}=p_0\left(\frac{M}{n}-1\right)a_{n-1} + [p_0(n-M+1)-r_0]a_n + r_0(n+1)a_{n+1}
\label{a_bulk}
\ee
for $n=1,\dots,M-2$ and 
\beqn 
\frac{da_0}{d\Gamma}= [p_0(1-M)-r_0]a_0 + r_0a_{1} \nonumber \\
\frac{da_{M-1}}{d\Gamma}=p_0\frac{1}{M-1}a_{M-2} -r_0a_{M-1}
\label{a_end}
\eeqn
for $n=0$ and $n=M-1$ ($M>1$).

\subsection{Critical point}

First, let us determine the unknown coefficient functions in the 
critical point. 
Using Eq. (\ref{prcrit}), we obtain a set of linear differential equations 
if $\gamma\equiv\ln(\Gamma+\Gamma_0)$ is used as an independent variable instead of $\Gamma$: 
\be 
\frac{da_n}{d\gamma}=\left(\frac{M}{n}-1\right)a_{n-1} + (n-M)a_n + (n+1)a_{n+1}
\ee
for $n=1,\dots,M-2$, and
\beqn
\frac{da_0}{d\gamma}=-Ma_0 + a_{1} \nonumber \\
\frac{da_{M-1}}{d\gamma}=\frac{1}{M-1}a_{M-2} - a_{M-1}
\eeqn
at the boundaries $n=0$, $n=M-1$ ($M>1$).
Arranging the functions in a column vector 
${\bf a}(\Gamma)=(a_{M-1}(\Gamma),a_{M-2}(\Gamma),\dots,a_0(\Gamma))^T$, 
the set of differential equations can be written in the form  
$\frac{d{\bf a}}{d\gamma}=A_M{\bf a}$, with a 
tridiagonal coefficient matrix 
\be 
A_M=\pmatrix{
-1   & \frac{1}{M-1}&               &        &               &    \cr
M-1  &-2            & \frac{2}{M-2} &        &               &    \cr
     & M-2          &  -3           & \frac{3}{M-3}&         &    \cr  
     &               &  M-3         & -4     & \ddots        &    \cr
     &               &               & \ddots & \ddots       & M-1\cr  
     &               &               &        &  1           & -M 
}.
\label{A}
\ee
The solution can then be expressed by the eigenvalues $-\epsilon_i$ and right eigenvectors ${\bf v}_i$ as   
\be 
{\bf a}(\Gamma)=\sum_{i=1}^{M}C_i{\bf v}_ie^{-\epsilon_i\gamma},
\label{aGamma}
\ee
where the coefficients $C_i$ are determined by the boundary condition, i.e. by specifying $s_{\Gamma}(\beta)$ at some $\Gamma$. 
Provided that the initial distribution of couplings at $\Gamma=0$ is the critical attractor given in Eq. (\ref{RP}), 
the initial condition is $s_{0}(\beta)=R_{0}(\beta)$, 
giving the initial value ${\bf a}(0)=(0,0,\dots,0,1)^T$ of the coefficient functions. 
The matrix $A_M$ is similar to a symmetric, tridiagonal matrix $-T_M=D_MA_MD_M^{-1}$, where $D_M$ is a diagonal matrix and $T_M$ is 
\be 
T_M=\pmatrix{
1   & 1&               &        &               &    \cr
1   &2            & \sqrt{2} &        &               &    \cr
    & \sqrt{2}          &  3           & \sqrt{3}&         &    \cr  
    &               &  \sqrt{3}         & 4     & \ddots        &    \cr
    &               &               & \ddots & \ddots       & \sqrt{M-1}\cr  
    &               &               &        &  \sqrt{M-1}          & M 
}.
\label{T}
\ee
This matrix, which has the eigenvalues $\epsilon_i$, is positive definite, as it must be in order to obtain an $S(\Gamma)$ that decreases with $\Gamma$\footnote{This can be shown by induction and by using the theorems that 
(i) a real symmetric tridiagonal matrix is positive definite if and only if its principal minors are positive and 
(ii) a real symmetric strictly diagonally dominant matrix (i.e. for which $|a_{ii}|>\sum_{j\neq i}|a_{ij}|$ for all $i$) with positive diagonal entries is positive definite.}.  
According to Eq. (\ref{aGamma}), the leading term in the sum for large $\gamma$ is the one containing the smallest eigenvalue $\epsilon_M$ of $T_M$, 
${\bf a}(\Gamma)\simeq C_M{\bf v}_Me^{-\epsilon_M\gamma}$. Then, using Eq. (\ref{Sgamma}), we arrive at an algebraic decrease of the survival probability of the central spin 
\be 
S(\Gamma)\sim\Gamma^{-\epsilon_M}
\ee
for large $\Gamma$.
This relation generalizes earlier results that have been known for the special cases $M=1$ and $M=2$ with $\epsilon_1=1$ and $\epsilon_2=\frac{3-\sqrt{5}}{2}$, respectively,  \cite{fisher,rm} to higher $M$, and provides a way of calculation of the $M$-dependent decay exponents. Numerical values of the latter for several $M$ can be found in Table \ref{table}.  
%%%%%%%%%%%%%%%%%%%%%%%%%%%%%%%%%%%%%%%%%%%%%%%%%%%%%%%%%%%%%%%%%%%
\begin{table}[h]
\begin{center}
\begin{tabular}{|r|r|r|}
\hline  $M$ & $\epsilon_M$ & $x_M$  \\
\hline
\hline 1&  $1$                       &  $1/2$  \\
\hline 2&  $3.8196601125\cdot 10^{-1}$&  $1.9098300563\cdot 10^{-1}$  \\
\hline 3&  $1.3919414689\cdot 10^{-1}$&  $6.9597073444\cdot 10^{-2}$  \\
\hline 4&  $4.3967261659\cdot 10^{-2}$&  $2.1983630830\cdot 10^{-2}$  \\
\hline 5&  $1.1448221145\cdot 10^{-2}$&  $5.7241105726\cdot 10^{-3}$  \\
\hline 6&  $2.4211175525\cdot 10^{-3}$&  $1.2105587763\cdot 10^{-3}$  \\
\hline 7&  $4.2197861966\cdot 10^{-4}$&  $2.1098930983\cdot 10^{-4}$  \\
\hline 8&  $6.222931429\cdot 10^{-5}$&   $3.111465714\cdot 10^{-5}$  \\
\hline 9&  $7.95732073\cdot 10^{-6}$&    $3.97866036\cdot 10^{-6}$  \\
\hline 10& $8.99215635\cdot 10^{-7}$&    $4.49607817\cdot 10^{-7}$  \\
\hline
\end{tabular}
\end{center}
\caption{\label{table} Numerical values of the decay exponent of the survival probability of the central spin obtained as the smallest eigenvalue of the matrix $T_M$ given in Eq. (\ref{T}) and the scaling dimension $x_M=\epsilon_M/2$ of the order parameter for different values of $M$.}
\end{table}
%%%%%%%%%%%%%%%%%%%%%%%%%%%%%%%%%%%%%%%%%%%%%%%%%%%%%%%%%%%%%%%%%%%%%%
Since $\epsilon_M$ is very small for large $M$, it is well approximated by
$\epsilon_M\approx -c^{(0)}_M/c^{(1)}_M$, where $c^{(0)}_M$ and $c^{(1)}_M$ are the zeroth and first order term, respectively, of the characteristic polynomial of $T_M$. 
The determinants $d_M=\det(T_M-\epsilon{\bf 1})$ for different $M$ can be shown to obey the recursion relation $d_M=(M-\epsilon)d_{M-1}-(M-1)d_{M-2}$, which results in the following recursions for $c^{(0)}_M$ and $c^{(1)}_M$: 
\beqn
c^{(0)}_M=Mc^{(0)}_{M-1}-(M-1)c^{(0)}_{M-2} \nonumber \\
c^{(1)}_M=Mc^{(1)}_{M-1}-(M-1)c^{(0)}_{M-2}-c^{(0)}_{M-1}.
\eeqn
With the initial conditions $c^{(0)}_1=c^{(0)}_2=1$ and 
$c^{(1)}_1=-1$, $c^{(1)}_2=-3$, we obtain that 
$c^{(0)}_M=1$ for all $M$, while $|c^{(1)}_M|\sim M!$ in leading order for large $M$. 
Thus, the smallest eigenvalue of $T_M$ approaches zero very rapidly, as 
\be 
\epsilon_M \sim 1/M!
\ee 
with increasing $M$.

\subsection{Griffiths-McCoy phase}

Let us now calculate the survival probability of the central spin in the paramagnetic Griffiths-McCoy phase, where the scale factors are given by Eq. (\ref{prgp}). 
Starting with the Eqs (\ref{a_bulk}-\ref{a_end}) and decomposing the coefficient functions in the form 
\be 
a_n(\Gamma)\equiv A_n(\Gamma)p_0(\Gamma),  \qquad n=0,1,\dots,M-1
\label{An}
\ee
we obtain the differential equations for the unknown functions $A_n(\Gamma)$
(for $M>1$)  
\beqn 
 \frac{dA_n}{d\Gamma}=p_0\left(\frac{M}{n}-1\right)A_{n-1} + p_0(n-M+1)A_n + r_0(n+1)A_{n+1} \quad (n=1,2,\dots,M-2) \nonumber \\
\frac{dA_0}{d\Gamma}= p_0(1-M)A_0 + r_0A_{1} \nonumber \\
\frac{dA_{M-1}}{d\Gamma}=p_0\frac{1}{M-1}A_{M-2}.
\label{diffA}
\eeqn
Taking into account that, according to Eq. (\ref{prgp}), $p_0(\Gamma)$ is vanishing compared to $r_0(\Gamma)$ in the limit $\Gamma\to\infty$, 
the leading order dependence of the functions $A_n(\Gamma)$ on $\Gamma$ can be obtained by neglecting the terms proportional to $p_0(\Gamma)$ on the r.h.s. of Eqs. (\ref{diffA}), resulting in
\beqn
 \frac{dA_n}{d\Gamma}\simeq \frac{1}{z}(n+1)A_{n+1} \quad (n=1,2,\dots,M-2) \nonumber \\
\frac{dA_0}{d\Gamma}\simeq\frac{1}{z}A_{1} \nonumber \\
\frac{dA_{M-1}}{d\Gamma}\simeq 0.
\label{diffAs}
\eeqn
This set of differential equations can be solved consecutively starting with $n=M-1$ for decreasing $n$, yielding 
$A_{M-1}(\Gamma)=C_{M-1}$, $A_{M-2}(\Gamma)=(M-1)\frac{1}{z}C_{M-1}\Gamma+C_{M-2}$, etc., where $C_n$, $n=0,1,\dots, M-1$ denote non-universal constants depending on the initial distribution of couplings. 
For a general index $n=M-l$, the function $A_{M-l}(\Gamma)$ will be  
a polynomial of $\Gamma$ of degree $l-1$: 
\be 
A_{M-l}(\Gamma)=\sum_{k=1}^lC_{M-k}{M-k \choose l-k}\left(\frac{\Gamma}{z}\right)^{l-k} \qquad l=1,2,\dots,M.
\ee
Specially, for $n=0$, we obtain 
$A_0(\Gamma)=\sum_{k=1}^MC_{M-k}\left(\frac{\Gamma}{z}\right)^{M-k}$.
The leading order term in the survival probability of the central spin for a general $M$ will thus be 
\be 
S(\Gamma)\sim \left(\frac{\Gamma}{z}\right)^{M-1}e^{-\Gamma/z}.
\ee

\section{Critical and subcritical behavior near the junction}
\label{scaling}

\subsection{Magnetization of the RTIM}

We will now discuss how the survival probability of the central spin calculated in the previous section is related to physical properties of the RTIM. 
Let us consider the average local magnetization $\overline{m}(H)=\overline{\langle\sigma_0^x\rangle}(H)$ in response to a small longitudinal magnetic field $H$ applied on the central spin. 
As it was argued in Ref. \cite{fisher}, the local magnetization at a given spin for a fixed $H$ is $O(1)$ in samples in that the spin remains active when the model is renormalized down to the energy scale $\Gamma_H\equiv\ln(\Omega_0/H)$. 
In samples in that the cluster containing the given spin had been decimated at some energy scale $\Omega$, the magnetization will be $O(\frac{H}{\Omega})$, which is vanishing as $H\to 0$.  
Therefore, in the critical point, the average magnetization at the central spin is given by the fraction of samples with an $O(1)$ magnetization, which is nothing but the survival probability of the central spin at the scale $\Gamma_H$: 
\be
\overline{m}(H)\sim S(\Gamma_H) \sim \left[\ln\left(\frac{\Omega_0}{H}\right)\right]^{-\epsilon_M}  \qquad (\Delta=0).
\ee
In the paramagnetic Griffiths-McCoy phase, but close enough to the critical point, so that  $z>1$, the average is still dominated by samples with an $O(1)$ magnetization, yielding
\be 
\overline{m}(H)\sim S(\Gamma_H) \sim \left[\ln\left(\frac{\Omega_0}{H}\right)\right]^{M-1}
\left(\frac{H}{\Omega_0}\right)^{1/z} \quad (z>1).
\ee
If, however, $z<1$, the average magnetization is determined by the contribution of typical samples, resulting in a linear dependence on $H$, $\overline{m}(H)\sim H$. 

We can see that the average spontaneous magnetization $\overline{m}_0=\lim_{H\to 0}\overline{\langle\sigma_0^x\rangle}(H)$ is zero in the critical point and below. It is, however, positive in the ferromagnetic phase and vanishes singularly at the transition point as  $\overline{m}_0(\Delta)\sim \Delta^{\beta_M}$. 
The order parameter exponent $\beta_M$ can be related to $\epsilon_M$ by scaling arguments as follows. 
In the ferromagnetic phase but close to the critical point ($\Delta\ll 1$), the correlation length $\xi_{\Delta}$ of average correlations is finite and, well beyond this length scale, $\xi\gg \xi_{\Delta}$, almost always bond decimations occur and field decimations not in the SDRG procedure. The corresponding characteristic energy scale is $\Gamma_{\Delta}\sim \xi_{\Delta}^{\psi}$. 
Since, close to the critical point, the correlation length diverges as $\xi_{\Delta}\sim \Delta^{-\nu}$ with $\nu=2$ \cite{fisher}, 
we obtain for the spontaneous magnetization 
\be  
\overline{m}_0(\Delta)\sim S(\Gamma_{\Delta})\sim \Gamma_{\Delta}^{-\epsilon_M}
\sim \xi_{\Delta}^{-\psi\epsilon_M}\sim \Delta^{\nu\psi\epsilon_M}.
\ee
Substituting the known values of the exponents $\nu=2$ and $\psi=1/2$, we obtain that the order parameter exponent is 
\be
\beta_M=\epsilon_M,
\ee
and the scaling dimension of the order parameter given by $x_M=\beta_M/\nu$ is $x_M=\epsilon_M/2$. 

\subsection{Dynamics in the random contact process}

A frequently studied quantity of the random contact process is the average probability 
$\overline{\rho_n}(t)$ that site $n$ is active at time $t$ if, initially, all sites were active. Owing to the duality property of the model, $\overline{\rho_n}(t)$ is identical to the average probability $\overline{\mathcal{P}_n(t)}$ that the process has not yet reached its absorbing (empty lattice) state at time $t$ provided that, initially, only site $n$ was active \cite{hv}. 

Far from the central site, $\overline{\mathcal{P}_n(t)}$ does not differ from that of the infinite chain, at least up to time $\ln t\sim l^{\psi}$, where $l$ is the distance from the center. But close to the junction, it is affected by the modified topology at the center, so we concentrate on the case when the initially active site is the central one ($n=0$). 
Then the function $\overline{\mathcal{P}_0(t)}$ is given by the probability that the central site is still not decimated at the scale $\Omega=t^{-1}$ in the SDRG procedure \cite{hiv,j1}, i.e. 
\be
\overline{\mathcal{P}_0(t)}\sim S[\ln(\Omega_0t)].
\ee
Using the results of section \ref{sdrg}, 
this leads to an asymptotic time-dependence  
\be 
\overline{\mathcal{P}_0(t)}\sim (\ln t)^{-\overline{\delta}_M}
\qquad (\Delta=\Delta_c),
\ee
with the decay exponent $\overline{\delta}_M=\epsilon_M$ in the critical point, while, in the inactive Griffiths-McCoy phase, we obtain an algebraic decay with a multiplicative polylogarithmic correction:  
\be 
\overline{\mathcal{P}_0(t)}\sim (\ln t)^{M-1}t^{-1/z}
\qquad (\Delta<\Delta_c).
\ee
The local order parameter of the transition 
$\overline{\mathcal{P}_0(\infty)}\equiv\lim_{t\to\infty}\overline{\mathcal{P}_0(t)}$ vanishes as the critical point is approached in the active phase similar to the 
RTIM: 
\be 
\overline{\mathcal{P}_0(\infty)} \sim (\Delta-\Delta_c)^{\beta_M}
\ee
with $\beta_M=\epsilon_M$.

\section{Discussion}
\label{discussion}

We have studied in this work models with an infinite-disorder critical behavior (the random transverse-field Ising model and the random contact process) at a star junction of $M$ chains. In the critical point, the scaling dimension $x_M$ of the local order parameter at the junction has been calculated by the SDRG method while, in the disordered Griffiths-McCoy phase, the scaling of the order parameter with the energy scale (or with the length scale) has been determined, generalizing thereby previous results for semi-infinite and infinite chains to multiple junctions ($M>2$) of chains. 
We have found that, although the scaling dimension $x_M$ decreases with the increasing number $M$ of arms very rapidly, it remains positive for any finite $M$. Thus, the local magnetization vanishes continuously at the critical point, no matter how large $M$ is. 
This behavior contrasts with that of the pure RTIM, where the transition becomes of first order above a threshold value $M=2$ \cite{itb}.  
Nevertheless, the tiny values of $x_M$ for not too small $M$ --- 
note that, already for $M=5$, it is $O(10^{-3})$ --- hardly differ from zero, which is formally characteristic of a first-order transition, 
making the transition practically indistinguishable from discontinuous ones by numerical methods. 

In the pure transverse-field Ising model, the extrapolation of $x_M$ to $M=0$ gives the magnetic exponent of the classical, two-dimensional Ising model in the presence of a random surface field \cite{itb}. Although, in our model, the chains are connected in a somewhat different way, this is expected to be an unimportant difference regarding the large scale behavior. Therefore, the extrapolation of $x_M$ to $M=0$ 
should give the surface magnetic exponent of the McCoy-Wu model (two-dimensional Ising model with columnar disorder) in the presence of a random surface field. 
But, as opposed to the pure variant, where $x_M$ is constant for $M<2$, 
$x_M$ varies with $M$ in the disordered model and, as it is known only for integers, it is not possible to extrapolate it to $M=0$ unambiguously. 

In this work, we have restricted ourselves to the study of disorder-averaged quantities. In order to obtain information on the distribution of sample-dependent quantities one must go beyond the decimation of the central spin, which is a difficult task. In general, the rescaled spontaneous magnetization 
$x=\ln(1/m_0)\frac{1}{z'}$, where $z'$ is the dynamical exponent in the ferromagnetic phase, is expected to have some limit distribution $f_M(x)$ as the critical point is approached ($1/z'\to 0$) \cite{j1}.
It has been pointed out that, in general, $\delta=\ln(1/\mathcal{P}_0(t)/\ln t)$ in the random contact process has the same limit distribution for $t\to\infty$ as $x$ in the RTIM \cite{j1}. 
The limit distribution $f_1(x)$ is known to be a pure exponential \cite{fisher} and although, for the infinite chain ($M=2$), the function $f_2(x)$ is unknown, 
numerical investigations have shown that it has an asymptotic form $e^{-2x}$ for large $x$ as if the chain were composed of two independent semi-infinite chains, and an asymptotic form  $f_2(x)\sim x^{-1+\overline{\delta_2}}$ for small $x$ \cite{j1}. 
Based on these earlier results, we conjecture that, for multiple junctions ($M>2$), the limit distribution decays for large $x$ as $f_M(x)\sim e^{-Mx}$ while, for small $x$, it diverges as $f_M(x)\sim x^{-1+\overline{\delta_M}}$.

\ack
%%%%%%%%%%%%%%%%%%%%%%%%%%%%%%%%%%%%%%%%%%%%%%%%%%%%%%%%%%%%%%%%%%%%%%
Discussions with F. Igl\'oi are gratefully acknowledged. 
This work was supported by the J\'anos Bolyai Research Scholarship of the
Hungarian Academy of Sciences, by the National Research Fund
under grant no. K75324, K109577, and partially supported by 
the European Union and the
European Social Fund through project FuturICT.hu (grant no.:
TAMOP-4.2.2.C-11/1/KONV-2012-0013).

%%%%%%%%%%%%%%%%%%%%%%%%%%%%%%%%%%%%%%%%%%%%%%%%%%%%%%%%%%%%%%%%%%%%%%%
%%%%%%%%%%%%%%%%%%%%%%%%%%%%%%%%%%%%%%%%%%%%%%%%%%%%%%%%%%%%%%%%%%%%%%%


\section*{References}
\begin{thebibliography}{99}




\bibitem{im}
Igl\'oi F, Monthus C 2005 Phys. Rep. {\bf 412} 277.

\bibitem{vojta_rev}
Vojta T 2006 J. Phys. A {\bf 39} R143.

\bibitem{fisher}
Fisher D S 1992 Phys. Rev. Lett. {\bf 69} 534;
1995 Phys. Rev. B {\bf 51} 6411.

\bibitem{mdh}
Ma S K, Dasgupta C, and Hu C K 1979 Phys. Rev. Lett. {\bf 43} 1434. 

\bibitem{cp} Harris T E  1974 Ann. Prob. {\bf 2} 969

\bibitem{liggett} Liggett T M 1999 {\it Stochastic interacting systems:
  contact, voter, and exclusion processes} (Berlin, Springer).

\bibitem{hhl} 
Henkel M, Hinrichsen H, L\"ubeck S 2008 
{\it Non-Equilibrium Phase transitions} Springer, Berlin 

\bibitem{odor}
\'Odor G 2008 {\it Universality in Nonequilibrium Lattice Systems} 
World Scientific, Singapore; 
2004 Rev. Mod. Phys. {\bf 76} 663

\bibitem{hiv} 
Hooyberghs J, Igl\'oi F, Vanderzande C 
2003 Phys. Rev. Lett. {\bf 90} 100601;
2004 Phys. Rev. E {\bf 69} 066140

\bibitem{vd} Vojta T and Dickison M 2005 Phys. Rev. E {\bf 72} 036126

\bibitem{ki_d}
Kov\'acs I A, Igl\'oi F 2010 Phys. Rev. B {\bf 82} 054437; 
2011 Phys. Rev. B {\bf 83} 174207

\bibitem{mccoy}
McCoy B M 1969 Phys. Rev. {\bf 188} 1014.

\bibitem{bjornberg}
Bj\"ornberg J E 2009 J. Stat. Phys. {\bf 135} 571

\bibitem{itb}
Igl\'oi F, Turban L, Berche B 1991 J. Phys. A: Math. Gen. {\bf 24} L1031


\bibitem{tsvelik}
Tsvelik A M 2013 Phys. Rev. Lett {\bf 110} 147202;
2014 New J. Phys. {\bf 16} 033003. 


\bibitem{griffiths} Griffiths R B 1969 Phys. Rev. Lett. {\bf 23} 17;
McCoy B M 1969 Phys. Rev. Lett. {\bf 23} 383

\bibitem{igloi}
Igl\'oi F 2002  Phys. Rev. B {\bf 65} 064416.

\bibitem{rm}
Refael G, Moore J E 2004 Phys. Rev. Lett. {\bf 93} 260602.

\bibitem{hv}
Hooyberghs J, Vanderzande C 2001 Phys. Rev. E  {\bf 63}, 041109

\bibitem{j1}
Juh\'asz R 2014 Phys. Rev. E {\bf 89} 032108.








\end{thebibliography}
\end{document}